\documentclass[aps,tightlines,superscriptaddress,groupedaddress,showkeys]{revtex4}
\usepackage{graphicx}
\usepackage{dcolumn}
\usepackage{bm}
\usepackage{amssymb}
\usepackage{amsfonts}
\usepackage{amsmath}

\begin{document}

\title{Upper bound on three-tangles of reduced states of four-qubit pure
states}
\author{S. Shelly Sharma}
\email{shelly@uel.br}
\affiliation{Departamento de Fisica, Universidade Estadual de Londrina, Londrina-PR,
Brazil}
\author{N. K. Sharma}
\email{nsharma@uel.br}
\affiliation{Departamento de Matematica, Universidade Estadual de Londrina, Londrina
86051-990, PR Brazil }
\keywords{Multipartite entanglement, Three-tangle, Four-qubit pure states,
Three-qubit invariants of pure state, Local unitary invariance}

\begin{abstract}
Closed formulae for upper bound on three-tangles of three-qubit reduced
states in terms of three-qubit invariant polynomials of pure four-qubit
states are obtained. Our results offer tighter constraints on total
three-way entanglement of a given qubit with the rest of the system than
those used in ref. \cite{regu14,regu16} \ to verify monogamy of four-qubit
quantum entanglement.
\end{abstract}

\maketitle

\section{Introduction}

Statistical mixtures of quantum states have quantum as well classical
correlations. Detecting and quantifying nonclassicality (or quantumness) of
multipartite quantum system is of special interest in quantum optics \cite%
{horo09, dodo03,peri94}, quantum information \cite{niel00}, and even in
quantum biology \cite{lamb13,li12}. In particular, detecting and quantifying
quantum entanglement is increasingly important to experiments in quantum
information science \cite{guhn07,eise07,wund09} and it also plays a crucial
role in investigations in quantum statistical physics, e.g., in studying
phase transitions \cite{sahl15,oste02}. State coefficients of a pure quantum
state contain complete information about the correlations amongst the
subsystems. Local unitary invariance constraints allow us to construct
unitary invariant functions of state coefficients to quantify two-way,
three-way, four-way,..., N-way entanglement of an N-qubit state. Global
negativity \cite{zycz98,vida02} or equivalently concurrence \cite{hill97} of
a pure state of two qubits quantifies two-way correlations. Three tangle 
\cite{coff00} is an entanglement monotone that quantifies three-way
entanglement of a three-qubit state. Four-way entanglement of a four-qubit
state is quantified by four-tangle \cite{shar13}, which for pure state is a
degree eight function of state coefficients. Three-tangle and four-tangle
for pure quantum states are moduli of specific local special unitary (LSUT)
invariants, that are polynomial functions of state coefficients.
Non-locality properties of an entangled state do not change under unitary
operations acting independently on each of its sub-systems. The idea of
describing entanglement by means of local-unitary invariants has been
explored in refs. \cite{gras98,luqu03,luqu06}. It is shown in ref. \cite%
{luqu06} that given a basis of stochastic local operations and classical
communication (SLOCC) covariants of degree $d$ in variables, the scalar
products with respect to auxilliary variables, form a basis of the space of
local unitary (LUT) invariants. Squared modulus $|I|^{2}$ of a SLOCC
invariant is a LUT and LSUT invariant.

Using an intuitive approach coupled to classical theory of invariants it has
been shown in ref. \cite{shar16} that one can construct local unitary
invariants characterizing a pure state of N two-level quantum systems
(qubits) in terms of N-1 qubit invariants. A natural question is, can we
write entanglement of (N-1)-qubit marginals of an N-qubit pure state in
terms of invariants of the N-qubit pure state? In this letter, we focus on
four qubit states and their three-qubit marginals. Entanglement measures for
reduced states are, generally, based on the convex roof of a quantity
defined on pure states. In most of the cases they are not computable as
there are no efficient ways to calculate convex roofs. Although a closed
formula for the three-tangle \cite{coff00} of pure states is known, for
mixed states no such formula is available. Solutions to finding the convex
roof of three-tangle for special families of states are, however, available
in \cite{lohm06,elts08,elts12,elts14}. A method to compute entanglement
measures based on convex roof constructions is also given in ref. \cite%
{toth15}. Recently, Osterloh \cite{oste16}, obtained three tangles of nine
classes of four qubit states \cite{vers02}, by finding a decomposition of
three qubit mixed state such that its entanglement lies on minimal
characteristic curve. We outline the formalism to construct three-tangle of
reduced state from relevant three-qubit polynomial invariants of the four
qubit pure state. Local unitary invariance of entanglement is the basic
principle underlying the construction. In the most general case, it involves
finding the roots of a quartic equation. The roots may be found analytically
or numerically. Closed formulae for upper bound on three tangles of nine
classes of four qubit states \cite{vers02}, are obtained and the results
compared with those of ref. \cite{regu14,regu16}. Our results offer tighter
constraints on three-way entanglement of a given qubit with the rest of the
system than those used in ref. \cite{regu14,regu16}.

We start by defining three tangle on pure and mixed states in section II,
outline the formalism in sections III and IV, discuss the upper bounds on
three tangles for nine classes of four-qubit states in section V and optimal
decomposition of a rank two mixed state in section VI. Section VII contains
the concluding remarks.

\section{Definition of Three Tangle}

Consider a three-qubit pure state%
\begin{equation}
\left\vert \Psi _{3}\right\rangle
=\sum\limits_{i_{1},i_{2},i_{3}}a_{i_{1}i_{2}i_{3}}\left\vert
i_{1}i_{2}i_{3}\right\rangle ,\quad i_{m}=0,1
\end{equation}%
where state coefficients $a_{i_{1}i_{2}i_{3}}$ are complex numbers and $%
i_{m}\ $refers to computational basis state of qubit $A_{m}$, $\left(
m=1,2,3\right) $. Let qubit $A_{1}$ be the focus qubit. Using the notation
from ref. \cite{shar16}, we define $D_{\left( A_{3}\right)
_{i_{3}}}^{00}=a_{00i_{3}}a_{11i_{3}}-a_{10i_{3}}a_{01i_{3}},$ ($i_{3}=0,1$)
as determinant of a two-way negativity font and $%
D^{00i_{3}}=a_{00i_{3}}a_{11i_{3}+1}-a_{10i_{3}}a_{01i_{3}+1}$, ($i_{3}=0,1$%
) is determinant of a three-way negativity font. Three tangle of pure state $%
\left\vert \Psi _{3}\right\rangle $ as defined in ref. \cite{coff00} is
equal to the modulus of a polynomial invariant of degree four that is $\tau
_{A_{1}|A_{2}|A_{3}}\left( \left\vert \Psi _{3}\right\rangle \right)
=4\left\vert I_{3,4}\right\vert $, where 
\begin{equation}
I_{3,4}=\left( D^{000}+D^{001}\right) ^{2}-4D_{\left( A_{3}\right)
_{0}}^{00}D_{\left( A_{3}\right) _{1}}^{00}.  \label{three_invariant}
\end{equation}%
The entanglement measure $\tau _{A_{1}|A_{2}|A_{3}}\left( \left\vert \Psi
_{3}\right\rangle \right) $ is extended to a mixed state of three qubits via
convex roof extension that is%
\begin{equation}
\left[ \tau _{A_{1}|A_{2}|A_{3}}\left( \rho _{3}\right) \right] ^{\frac{1}{2}%
}=\min_{\left\{ p_{i},\left\vert \phi ^{\left( i\right) }\right\rangle
\right\} }\sum\limits_{i}p_{i}\left[ \tau _{A_{1}|A_{2}|A_{3}}\left(
\left\vert \phi ^{\left( i\right) }\right\rangle \right) \right] ^{\frac{1}{2%
}},  \label{mix3tangle}
\end{equation}%
where minimization is taken over all possible decompositions $\left\{
p_{i},\left\vert \phi ^{\left( i\right) }\right\rangle \right\} $ of $\rho
_{3}$.

\section{Upper bound on three tangle of a rank two reduced state}

Purification of a rank two three-qubit mixed state is a four-qubit pure
state. In this section, we describe the construction of upper bound on three
tangle of marginal state of a four qubit pure state in terms of three-qubit
and four-qubit unitary invariants of the pure state. In the most general
case, it involves finding the roots of a quartic equation.The roots may be
found analytically or numerically. Analytical form of relevant three-qubit
and four-qubit invariants is to be found in Sec. V of ref. \cite{shar16}.
Upper bounds on three-qubit mixed state entanglement, used in ref. \cite%
{regu14,regu16} \ in the context of a monogamy of quantum entanglement, had
been calculated by using an algorithm \cite{rodr14} to construct the
entanglement-minimizing decomposition for $\rho $. Our results offer tighter
constraints on three-way entanglement of a given qubit with the rest of the
system than those used in ref. \cite{regu14,regu16}.

Consider a general four qubit pure state, written as

\begin{equation}
\left\vert \Psi \right\rangle =\sum_{i_{1},i_{2},i_{3}}\left(
a_{i_{1}i_{2}i_{3}0}\left\vert i_{1}i_{2}i_{3}0\right\rangle
+a_{i_{1}i_{2}i_{3}1}\left\vert i_{1}i_{2}i_{3}1\right\rangle \right) ,\quad
\left( i_{m}=0,1\right) ,  \label{4state}
\end{equation}%
where qubits are in the order ($A_{1}$, $A_{2}$, $A_{3}$, $A_{4}$)
corresponding to the state $\left\vert i_{1}i_{2}i_{3}i_{4}\right\rangle $.
Choosing qubit $A_{1}$ as the focus qubit, we identify $D_{\left(
A_{3}\right) _{i_{3}}\left( A_{4}\right)
_{i_{4}}}^{00}=a_{00i_{3}i_{4}}a_{11i_{3}i_{4}}-a_{10i_{3}i_{4}}a_{01i_{3}i_{4}}
$ (two-way), $D_{\left( A_{2}\right) _{i_{2}}\left( A_{4}\right)
_{i_{4}}}^{00}=a_{0i_{2}0i_{4}}a_{1i_{2}1i_{4}}-a_{1i_{2}0i_{4}}a_{0i_{2}1i_{4}}
$ (two-way), $D_{\left( A_{2}\right) _{i_{2}}\left( A_{3}\right)
_{i_{3}}}^{00}=a_{0i_{2}i_{3}0}a_{1i_{2}i_{3}1}-a_{1i_{2}i_{3}0}a_{0i_{2}i_{3}1}
$ (two-way), $D_{\left( A_{4}\right)
_{i_{4}}}^{00i_{3}}=a_{00i_{3}i_{4}}a_{11,i_{3}\oplus
1,i_{4}}-a_{10i_{3}i_{4}}a_{01,i_{3}\oplus 1,i_{4}}$ (three-way), $D_{\left(
A_{3}\right) _{i_{3}}}^{00i_{4}}=a_{00i_{3}i_{4}}a_{11i_{3},i_{4}\oplus
1}-a_{10i_{3}i_{4}}a_{01i_{3},i_{4}\oplus 1}$ (three-way), $D_{\left(
A_{2}\right) _{i_{2}}}^{00i_{4}}=a_{0i_{2}0i_{4}}a_{1i_{2}1i_{4}\oplus
1}-a_{1i_{2}0i_{4}}a_{0i_{2}1i_{4}\oplus 1}$ (three-way), and $%
D^{00i_{3}i_{4}}=a_{00i_{3}i_{4}}a_{11,i_{3}\oplus 1,i_{4}\oplus
1}-a_{10i_{3}i_{4}}a_{01,i_{3}\oplus 1,i_{4}\oplus 1}$- (four-way) as the
determinants of negativity fonts. We shall also use the three and four qubit
invariants constructed in section V of ref. \cite{shar16}. Three-qubit
invariants of interest for the triple $A_{1}A_{2}A_{3}$ in state $\left\vert
\Psi \right\rangle $ comprise a set denoted by $\left\{ \left(
I_{3,4}\right) _{A_{4}}^{4-m,m}:m=0\text{ to }4\right\} $. The form of
elements of the set in terms of determinants of negativity fonts is given in
Appendix \ref{A1}. The elements in set $\left\{ \left( I_{3,4}\right)
_{A_{4}}^{4-m,m}:m=0\text{ to }4\right\} $ are invariant with respect to
local unitaries on qubits $A_{1}$, $A_{2}$, and $A_{3}$. Four-qubit
invariant that quantifies the sum of three-way and four-way correlations 
\cite{shar16} of triple $A_{1}A_{2}A_{3},$ reads as%
\begin{equation}
N_{4,8}^{A_{1}A_{2}A_{3}}=6\left\vert \left( I_{3,4}\right)
_{A_{4}}^{2,2}\right\vert ^{2}+4\left\vert \left( I_{3,4}\right)
_{A_{4}}^{3,1}\right\vert ^{2}+4\left\vert \left( I_{3,4}\right)
_{A_{4}}^{1,3}\right\vert ^{2}+\left\vert \left( I_{3,4}\right)
_{A_{4}}^{4,0}\right\vert ^{2}+\left\vert \left( I_{3,4}\right)
_{A_{4}}^{0,4}\right\vert ^{2},
\end{equation}%
while degree eight invariant that detects genuine four-body entanglement of
a four-qubit state is given by%
\begin{equation}
I_{4,8}=3\left( \left( I_{3,4}\right) _{A_{4}}^{2,2}\right) ^{2}-4\left(
I_{3,4}\right) _{A_{4}}^{3,1}\left( I_{3,4}\right) _{A_{4}}^{1,3}+\left(
I_{3,4}\right) _{A_{4}}^{4,0}\left( I_{3,4}\right) _{A_{4}}^{0,4}.
\label{i4inv}
\end{equation}%
Invariant $I_{4,8}$ is written as a function of $A_{1}A_{2}A_{3}$ invariants
in Eq. (\ref{i4inv}). Being independent of the choice of focus qubit, $%
I_{4,8}$ can be written in alternate forms. Four-tangle based on degree
eight invariant is defined \cite{shar13,shar16} as%
\begin{equation*}
\tau _{4,8}=16\left\vert 12\left( I_{4,8}\right) \right\vert .
\end{equation*}%
Set$\left\{ \left( I_{3,4}\right) _{A_{3}}^{4-m,m}:m=0\text{ to }4\right\} $
for the triple $A_{1}A_{2}A_{4}$ and $\left\{ \left( I_{3,4}\right)
_{A_{2}}^{4-m,m}:m=0\text{ to }4\right\} $ for qubits $A_{1}A_{3}A_{4}$ can
be constructed from two-qubit invariants of properly selected pair of qubits 
\cite{shar16}. In the following sections, the subscript $4$ in $\left(
I_{3,4}\right) _{A_{3}}^{4-m,m}$ has been dropped, that is $\left(
I_{3,4}\right) _{A_{3}}^{4-m,m}=\left( I_{3}\right) _{A_{3}}^{4-m,m}$ .

\subsection{Unitary on fourth qubit}

To illustrate the method, we focus on three-qubit reduced state $\rho
=\sum_{i=0}^{1}p_{A_{4}}^{\left( i\right) }\left\vert \phi _{A_{4}}^{\left(
i\right) }\right\rangle \left\langle \phi _{A_{4}}^{\left( i\right)
}\right\vert $, obtained by tracing out qubit $A_{4}$\ from the state $%
\left\vert \Psi \right\rangle $. Probability of finding qubit $A_{4}$ in
state $\left\vert i\right\rangle $ is denoted by $p_{A_{4}}^{\left( i\right)
}$. A unitary $U(x)=\frac{1}{\sqrt{1+\left\vert x\right\vert ^{2}}}\left[ 
\begin{array}{cc}
1 & -x^{\ast } \\ 
x & 1%
\end{array}%
\right] $, on qubit $A_{4}$ of state $\left\vert \Psi \right\rangle $
results in an$\ x$ dependent state%
\begin{equation}
\left\vert \Psi \left( x\right) \right\rangle =\sum_{i_{4}=0}^{1}\left\vert
\Phi _{A_{4}}^{\left( i_{4}\right) }(x)\right\rangle \left\vert
i_{4}\right\rangle ,
\end{equation}%
where $\left\vert \Phi _{A_{4}}^{\left( i_{4}\right) }(x)\right\rangle
=\sum_{i_{1},i_{2},i_{3}}a_{i_{1}i_{2}i_{3}i_{4}}(x)\left\vert
i_{1}i_{2}i_{3}\right\rangle $ is a subnormalized state. Reduced state
obtained by tracing out qubit $A_{4}$\ reads as%
\begin{equation}
\rho (x)=\sum_{i=0}^{1}p_{A_{4}}^{\left( i\right) }(x)\left\vert \phi
_{A_{4}}^{\left( i\right) }(x)\right\rangle \left\langle \phi
_{A_{4}}^{\left( i\right) }(x)\right\vert ,
\end{equation}%
where $\left\vert \phi _{A_{4}}^{\left( i_{4}\right) }(x)\right\rangle =%
\frac{\left\vert \Phi _{A_{4}}^{\left( i_{4}\right) }(x)\right\rangle }{%
\sqrt{p_{A_{4}}^{\left( i\right) }(x)}}$, is a normalized state, and $x$
dependent probabilities are defined as 
\begin{equation}
p_{A_{4}}^{\left( 0\right) }(x)=\frac{p_{A_{4}}^{\left( 0\right)
}+\left\vert x\right\vert ^{2}p_{A_{4}}^{\left( 1\right) }}{1+\left\vert
x\right\vert ^{2}},\hspace{0.3in}p_{A_{4}}^{\left( 1\right) }(x)=\frac{%
p_{A_{4}}^{\left( 1\right) }+\left\vert x\right\vert ^{2}p_{A_{4}}^{\left(
0\right) }}{1+\left\vert x\right\vert ^{2}}.
\end{equation}%
One can verify that the $x$ dependent three-tangle%
\begin{equation}
\tau _{A_{1}|A_{2}|A_{3}}\left( \left\vert \phi _{A_{4}}^{\left( 0\right)
}(x)\right\rangle \right) =\frac{4}{\left( p_{A_{4}}^{\left( 0\right)
}(x)\right) ^{2}}\left\vert \left( I_{3}\right) _{A_{4}}^{4,0}\left(
x\right) \right\vert ,
\end{equation}%
and 
\begin{equation}
\tau _{A_{1}|A_{2}|A_{3}}\left( \left\vert \phi _{A_{4}}^{\left( 1\right)
}(x)\right\rangle \right) =\frac{4}{\left( p_{A_{4}}^{\left( 1\right)
}(x)\right) ^{2}}\left\vert \left( I_{3}\right) _{A_{4}}^{0,4}\left(
x\right) \right\vert .
\end{equation}%
Using the definition of three tangle for mixed states (Eq. (\ref{mix3tangle}%
)), we can write%
\begin{equation}
\left[ \tau _{A_{1}|A_{2}|A_{3}}\left( \rho _{3}\right) \right] ^{\frac{1}{2}%
}\leq 2\min_{\left\{ x\right\} }\left( \left\vert \left( I_{3}\right)
_{A_{4}}^{4,0}\left( x\right) \right\vert ^{\frac{1}{2}}+\left\vert \left(
I_{3}\right) _{A_{4}}^{0,4}\left( x\right) \right\vert ^{\frac{1}{2}}\right)
.  \label{ineq}
\end{equation}%
Three qubit invariants $\left( I_{3}\right) _{A_{4}}^{4,0}\left( x\right) $
and $\left( I_{3}\right) _{A_{4}}^{0,4}(x)$ are related to elements of the
set $\{\left( I_{3}\right) _{A_{4}}^{4-m,m}:m=0,4$\} through 
\begin{eqnarray}
\left( I_{3}\right) _{A_{4}}^{4,0}(x) &=&\frac{1}{\left( 1+\left\vert
x\right\vert ^{2}\right) ^{2}}\left[ \left( I_{3}\right)
_{A_{4}}^{4,0}-4x^{\ast }\left( I_{3}\right) _{A_{4}}^{3,1}\right.   \notag
\\
&&\left. +6\left( x^{\ast }\right) ^{2}\left( I_{3}\right) ^{2,2}-4\left(
x^{\ast }\right) ^{3}\left( I_{3}\right) _{A_{4}}^{1,3}+\left( x^{\ast
}\right) ^{4}\left( I_{3}\right) _{A_{4}}^{0,4}\right] ,  \label{i40}
\end{eqnarray}%
and%
\begin{eqnarray}
\left( I_{3}\right) _{A_{4}}^{0,4}(x) &=&\frac{1}{\left( 1+\left\vert
x\right\vert ^{2}\right) ^{2}}\left[ \left( I_{3}\right)
_{A_{4}}^{0,4}+4x\left( I_{3}\right) _{A_{4}}^{1,3}\right.   \notag \\
&&\left. +6x^{2}\left( I_{3}\right) _{A_{4}}^{2,2}+4x^{3}\left( I_{3}\right)
_{A_{4}}^{3,1}+\left( I_{3}\right) _{A_{4}}^{4,0}x^{4}\right] .  \label{i04}
\end{eqnarray}%
To obtain an upper bound on three tangle of mixed state, we find the value
of complex parameter $x_{1}$ such that $\left( I_{3}\right)
_{A_{4}}^{4,0}\left( x_{1}\right) =0$, and $x_{2}$ such that $\left(
I_{3}\right) _{A_{4}}^{0,4}\left( x_{2}\right) =0$. In the most general
case, that involves solving a quartic equation in variable $x$, (obtained
from Eq. (\ref{i40}) or (\ref{i04}) ). When the state coefficients are
known, the resulting quartic may be solved numerically. However, for the
representatives of nine classes of four-qubit entanglement \cite{vers02}
analytic solutions are easily found. By definition the three tangle must
satisfy Eq. (\ref{ineq}), as such three tangle satisfies the inequality%
\begin{equation}
\left[ \tau _{A_{1}|A_{2}|A_{3}}\left( \rho _{3}\right) \right] \leq 4\min
\left( \left\vert \left( I_{3}\right) _{A_{4}}^{0,4}\left( x_{1}\right)
\right\vert ,\left\vert \left( I_{3}\right) _{A_{4}}^{4,0}\left(
x_{2}\right) \right\vert \right) 
\end{equation}%
giving us an upper bound on three tangle of the mixed state.

\subsection{Unitary on three qubit state}

If normalized three-qubit states $\left\vert \phi _{A_{4}}^{\left( 0\right)
}\right\rangle $ and $\left\vert \phi _{A_{4}}^{\left( 1\right)
}\right\rangle $ are orthogonal to each other then a unitary on three qubit
state such that 
\begin{equation}
\left\vert \phi _{A_{4}}^{\left( 0\right) }(y)\right\rangle =\frac{1}{\sqrt{%
1+\left\vert y\right\vert ^{2}}}\left( \left\vert \phi _{A_{4}}^{\left(
0\right) }\right\rangle +y\left\vert \phi _{A_{4}}^{\left( 1\right)
}\right\rangle \right) ,
\end{equation}%
and 
\begin{equation}
\left\vert \phi _{A_{4}}^{\left( 1\right) }(y)\right\rangle =\frac{1}{\sqrt{%
1+\left\vert y\right\vert ^{2}}}\left( \left\vert \phi _{A_{4}}^{\left(
1\right) }\right\rangle -y^{\ast }\left\vert \phi _{A_{4}}^{\left( 0\right)
}\right\rangle \right) ,
\end{equation}%
results in a $y$ dependent four-qubit state%
\begin{equation}
\left\vert \Psi \left( y\right) \right\rangle =\sum_{i_{4}=0}^{1}\sqrt{%
p_{A_{4}}^{\left( i_{4}\right) }}\left\vert \phi _{A_{4}}^{\left(
i_{4}\right) }(y)\right\rangle \left\vert i_{4}\right\rangle ,
\end{equation}%
such that three tangle may be defined as%
\begin{equation}
\left[ \tau _{A_{1}|A_{2}|A_{3}}\left( \rho \right) \right] ^{\frac{1}{2}%
}=2\min_{\left\{ y\right\} }\left( p_{A_{4}}^{\left( 0\right) }\left\vert
\left( I_{3}\right) _{A_{4}}^{4,0}\left( y\right) \right\vert ^{\frac{1}{2}%
}+p_{A_{4}}^{\left( 1\right) }\left\vert \left( I_{3}\right)
_{A_{4}}^{0,4}\left( y\right) \right\vert ^{\frac{1}{2}}\right) .
\end{equation}%
Recalling that three-qubit invariants, $\left( I_{3}\right) _{A_{4}}^{k-m,m},
$ are degree four functions of state coefficients, in this case%
\begin{eqnarray}
\left( I_{3}\right) _{A_{4}}^{4,0}\left( y\right)  &=&\frac{1}{\left(
1+\left\vert y\right\vert ^{2}\right) ^{2}}\left( \frac{\left( I_{3}\right)
_{A_{4}}^{4,0}}{\left( p_{A_{4}}^{0}\right) ^{2}}+4y\frac{\left(
I_{3}\right) _{A_{4}}^{3,1}}{\sqrt{\left( p_{A_{4}}^{0}\right)
^{3}p_{A_{4}}^{1}}}\right.   \notag \\
&&\left. +6y^{2}\frac{\left( I_{3}\right) _{A_{4}}^{2,2}}{%
p_{A_{4}}^{0}p_{A_{4}}^{1}}+4y^{3}\frac{\left( I_{3}\right) _{A_{4}}^{1,3}}{%
\sqrt{p_{A_{4}}^{0}\left( p_{A_{4}}^{1}\right) ^{3}}}+y^{4}\frac{\left(
I_{3}\right) _{A_{4}}^{0,4}}{\left( p_{A_{4}}^{1}\right) ^{2}}\right) ,
\end{eqnarray}%
and 
\begin{eqnarray}
\left( I_{3}\right) _{A_{4}}^{0,4}\left( y\right)  &=&\frac{1}{\left(
1+\left\vert y\right\vert ^{2}\right) ^{2}}\left( \frac{\left( I_{3}\right)
_{A_{4}}^{04}}{\left( p_{A_{4}}^{1}\right) ^{2}}-4y^{\ast }\frac{\left(
I_{3}\right) _{A_{4}}^{1,3}}{\sqrt{\left( p_{A_{4}}^{1}\right)
^{3}p_{A_{4}}^{0}}}\right.   \notag \\
&&\left. +6\left( y^{\ast }\right) ^{2}\frac{\left( I_{3}\right)
_{A_{4}}^{2,2}}{p_{A_{4}}^{0}p_{A_{4}}^{1}}-4\left( y^{\ast }\right) ^{3}%
\frac{\left( I_{3}\right) _{A_{4}}^{3,1}}{\sqrt{p_{A_{4}}^{1}\left(
p_{A_{4}}^{0}\right) ^{3}}}+\left( y^{\ast }\right) ^{4}\frac{\left(
I_{3}\right) _{A_{4}}^{4,0}}{\left( p_{A_{4}}^{0}\right) ^{2}}\right) .
\end{eqnarray}%
We look for $\rho (y_{1})$ with $\left( I_{3}\right) _{A_{4}}^{4,0}\left(
y_{1}\right) =0$, and $\rho (y_{2})$ such that $\left( I_{3}\right)
_{A_{4}}^{0,4}\left( y_{2}\right) =0$. That gives us an upper bound on $%
\left[ \tau _{A_{1}|A_{2}|A_{3}}\left( \rho \right) \right] ^{\frac{1}{2}}$
that is 
\begin{equation}
\left[ \tau _{A_{1}|A_{2}|A_{3}}\left( \rho \right) \right] ^{\frac{1}{2}%
}\leq \min \left( 2p_{A_{4}}^{0}\left\vert \left( I_{3}\right)
_{A_{4}}^{0,4}\left( y_{1}\right) \right\vert ^{\frac{1}{2}}\text{, }%
2p_{A_{4}}^{1}\left\vert \left( I_{3}\right) _{A_{4}}^{4,0}\left(
y_{2}\right) \right\vert ^{\frac{1}{2}}\right) .  \label{tauup2}
\end{equation}

If for a given state $p_{A_{4}}^{\left( 0\right) }=p_{A_{4}}^{\left(
1\right) }$, then the result obtained by a unitary on fourth qubit coincides
with that obtained by a unitary on the three-qubit pure states of the
decomposition of the mixed state. When $p_{A_{4}}^{\left( 0\right) }\neq
p_{A_{4}}^{\left( 1\right) }$, then minima found by unitary on fourth qubit
and those calculated by unitary on three qubit state must be compared to
find the correct bound on three-tangle for the mixed state. Our results
complement the upper bounds of ref. \cite{oste16} which possibly correspond
to those found by applying a unitary on a three qubit marginal state (Eq. %
\ref{tauup2}).

\section{Three tangle and three-qubit invariants of six groups of states}

First of all we notice that since the difference $16\left(
N_{4,8}^{A_{i}A_{j}A_{k}}-2\left\vert I_{4,8}\right\vert \right) $ is a
measure of three-way correlations amongst qubits $A_{i}$, $A_{j}$, and $A_{k}
$ in pure state, the three tangle must satisfy the condition $\tau
_{A_{i}|A_{j}|A_{k}}\left( \rho \right) \leq 4\sqrt{%
N_{4,8}^{A_{i}A_{j}A_{k}}-2\left\vert I_{4,8}\right\vert }$. Evaluation of
three-qubit invariants $\left\{ \left( I_{3}\right) _{A_{q}}^{4-m,m}:m=0%
\text{ to }4\right\} $, where $A_{q}$ refers to the qubit which is traced
out, shows that three-qubit marginals of states representing nine classes of
four qubits belong in six groups. We use unitary on fourth qubit to express
upper bound on three tangle in terms of three-qubit invariants for the
following cases of interest:

(i) For a given triple of qubits $A_{i}A_{j}A_{k}$, $%
N_{4,8}^{A_{i}A_{j}A_{k}}=2\left\vert I_{4,8}\right\vert $, therefore three
tangle, $\tau _{A_{i}|A_{j}|A_{k}}\left( \rho \right) =0$.

(ii) Only $\left( I_{3}\right) _{A_{q}}^{4,0}$ is non-zero, therefore, $%
\left\vert I_{4,8}\right\vert =0$ and $\tau _{A_{i}|A_{j}|A_{k}}\left( \rho
\right) \leq 4\left\vert \left( I_{3}\right) _{A_{q}}^{4,0}\right\vert $.

(iii) Only $\left( I_{3}\right) _{A_{q}}^{0,4}$ \ is non-zero, therefore, $%
\left\vert I_{4,8}\right\vert =0$ and $\tau _{A_{i}|A_{j}|A_{k}}\left( \rho
\right) \leq 4\left\vert \left( I_{3}\right) _{A_{q}}^{0,4}\right\vert $.

(iv) Non zero three-qubit invariants are $\left( I_{3}\right) _{A_{q}}^{4,0}$%
and $\left( I_{3}\right) _{A_{q}}^{2,2},$ therefore%
\begin{equation}
\left( I_{3}\right) _{A_{q}}^{4,0}(x)=\frac{1}{\left( 1+\left\vert
x\right\vert ^{2}\right) ^{2}}\left( \left( I_{3}\right)
_{A_{q}}^{4,0}+6\left( x^{\ast }\right) ^{2}\left( I_{3}\right)
_{A_{q}}^{2,2}\right) ,
\end{equation}%
and%
\begin{equation}
\left( I_{3}\right) _{A_{q}}^{0,4}(x)=\frac{x^{2}}{\left( 1+\left\vert
x\right\vert ^{2}\right) ^{2}}\left( 6\left( I_{3}\right)
_{A_{q}}^{2,2}+x^{2}\left( I_{3}\right) _{A_{q}}^{4,0}\right) .
\end{equation}%
In this case three tangle satisfies the condition%
\begin{equation}
\tau _{A_{i}|A_{j}|A_{k}}\left( \rho \right) \leq 4\left\vert \left(
I_{3}\right) _{A_{4}}^{4,0}\right\vert \frac{\left\vert \left\vert 6\left(
I_{3}\right) ^{2,2}\right\vert -\left\vert \left( I_{3}\right)
_{A_{4}}^{4,0}\right\vert \right\vert }{\left\vert 6\left( I_{3}\right)
^{2,2}\right\vert +\left\vert \left( I_{3}\right) _{A_{4}}^{4,0}\right\vert }%
.  \label{tauup4}
\end{equation}

(v) Non zero three-qubit invariants are $\left( I_{3}\right) _{A_{q}}^{0,4}$
and $\left( I_{3}\right) _{A_{q}}^{2,2}$, then to obtain three tangle we use
the relations%
\begin{equation}
\left( I_{3}\right) _{A_{q}}^{4,0}(x)=\frac{\left( x^{\ast }\right) ^{2}}{%
\left( 1+\left\vert x\right\vert ^{2}\right) ^{2}}\left( 6\left(
I_{3}\right) _{A_{q}}^{2,2}+\left( x^{\ast }\right) ^{2}\left( I_{3}\right)
_{A_{q}}^{0,4}\right) ,
\end{equation}%
and%
\begin{equation}
\left( I_{3}\right) _{A_{q}}^{0,4}(x)=\frac{1}{\left( 1+\left\vert
x\right\vert ^{2}\right) ^{2}}\left( \left( I_{3}\right)
_{A_{q}}^{0,4}+x^{2}6\left( I_{3}\right) _{A_{q}}^{2,2}\right) ,
\end{equation}%
leading to the condition 
\begin{equation}
\tau _{A_{i}|A_{j}|A_{k}}\left( \rho \right) \leq 4\left\vert \left(
I_{3}\right) _{A_{4}}^{0,4}\right\vert \frac{\left\vert \left\vert 6\left(
I_{3}\right) ^{2,2}\right\vert -\left\vert \left( I_{3}\right)
_{A_{4}}^{0,4}\right\vert \right\vert }{\left\vert 6\left( I_{3}\right)
^{2,2}\right\vert +\left\vert \left( I_{3}\right) _{A_{4}}^{0,4}\right\vert }%
.  \label{tauup5}
\end{equation}

(vi) The special case where only $\left( I_{3}\right) _{A_{q}}^{0,4}$ and $%
\left( I_{3}\right) _{A_{q}}^{1,3}$ are non-zero such that%
\begin{equation}
\left( I_{3}\right) _{A_{q}}^{4,0}(x)=\frac{\left( x^{\ast }\right) ^{3}}{%
\left( 1+\left\vert x\right\vert ^{2}\right) ^{2}}\left( x^{\ast }\left(
I_{3}\right) _{A_{q}}^{0,4}-4\left( I_{3}\right) _{A_{q}}^{1,3}\right) ,
\end{equation}%
\begin{equation}
\left( I_{3}\right) _{A_{q}}^{0,4}(x)=\frac{1}{\left( 1+\left\vert
x\right\vert ^{2}\right) ^{2}}\left( \left( I_{3}\right)
_{A_{q}}^{0,4}+4x\left( I_{3}\right) _{A_{q}}^{1,3}\right) ,
\end{equation}%
therefore three tangle satisfies the inequality 
\begin{equation}
\tau _{A_{i}|A_{j}|A_{k}}\left( \rho \right) \leq \frac{4\left\vert \left(
I_{3}\right) _{A_{q}}^{0,4}\right\vert ^{3}}{\left\vert 4\left( I_{3}\right)
_{A_{q}}^{1,3}\right\vert ^{2}+\left\vert \left( I_{3}\right)
_{A_{q}}^{0,4}\right\vert ^{2}}.  \label{tauup6}
\end{equation}

\section{Three-tangles for nine classes of four-qubit entanglement}

In this section, we use the results from previous section to write down
upper bounds on three-tangles for representatives of nine classes of four
qubit states. Our results offer tighter constraints on total three-way
entanglement of a given qubit with the rest of the system than those used in
refs. \cite{regu14,regu16}.

\subsection{Class I}

Class one states are represented by

\begin{eqnarray}
\left\vert G_{abcd}^{(1)}\right\rangle &=&\frac{a+d}{2}\left( \left\vert
0000\right\rangle +\left\vert 1111\right\rangle \right) +\frac{a-d}{2}\left(
\left\vert 0011\right\rangle +\left\vert 1100\right\rangle \right)  \notag \\
&&+\frac{b+c}{2}\left( \left\vert 0101\right\rangle +\left\vert
1010\right\rangle \right) +\frac{b-c}{2}\left( \left\vert 0110\right\rangle
+\left\vert 1001\right\rangle \right) .
\end{eqnarray}%
For any partition $A_{i}A_{j}A_{k}$, $\left( I_{3}\right)
_{A_{l}}^{4,0}=\left( I_{3}\right) _{A_{l}}^{0,4}$, $\left( I_{3}\right)
_{A_{l}}^{2,2}\neq 0$, while $\left( I_{3}\right) _{A_{l}}^{3,1}=\left(
I_{3}\right) _{A_{l}}^{1,3}=0$. As a result, 16$N_{4,8}^{A_{i}A_{j}A_{k}}=2%
\left\vert I_{4,8}\right\vert $, therefore three tangle, $\tau
_{A_{i}|A_{j}|A_{k}}\left( \rho \right) =0$.

\subsection{Class II}

For class two states three-tangle for all four three-qubit partitions has
the same value. Consider pure state three-qubit invariants for partition $%
A_{1}A_{2}A_{3}$ in representative state%
\begin{eqnarray}
\left\vert G_{adc}^{(2)}\right\rangle  &=&\frac{a+d}{2}\left( \left\vert
0000\right\rangle +\left\vert 1111\right\rangle \right) +\frac{a-d}{2}\left(
\left\vert 0011\right\rangle +\left\vert 1100\right\rangle \right)   \notag
\\
&&+c\left( \left\vert 0101\right\rangle +\left\vert 1010\right\rangle
\right) +\left\vert 0110\right\rangle .
\end{eqnarray}%
Three-qubit invariants have values 
\begin{equation}
\left( I_{3}\right) _{A_{4}}^{4,0}=\frac{c\left( a^{2}-d^{2}\right) }{\left(
\left\vert a\right\vert ^{2}+\left\vert d\right\vert ^{2}+2\left\vert
c\right\vert ^{2}+1\right) ^{2}},\left( I_{3}\right) _{A_{4}}^{2,2}=\frac{%
\left( a^{2}-c^{2}\right) \left( d^{2}-c^{2}\right) }{6\left( \left\vert
a\right\vert ^{2}+\left\vert d\right\vert ^{2}+2\left\vert c\right\vert
^{2}+1\right) ^{2}},
\end{equation}%
while $\left( I_{3}\right) _{A_{4}}^{3,1}=\left( I_{3}\right)
_{A_{4}}^{1,3}=\left( I_{3}\right) _{A_{4}}^{0,4}=0$. Consequently the sum
of three and four-way correlations is given by 
\begin{equation}
16N_{4,8}^{A_{1}A_{2}A_{3}}=16\left( \left\vert \left( I_{3}\right)
_{A_{4}}^{4,0}\right\vert ^{2}+6\left\vert \left( I_{3}\right)
_{A_{4}}^{2,2}\right\vert ^{2}\right) .
\end{equation}%
Using the result of Eq. (\ref{tauup4}) and the fact that $\tau
_{A_{i}|A_{j}|A_{k}}\left( \rho \right) =\tau _{A_{1}|A_{2}|A_{3}}\left(
\rho \right) $, we obtain 
\begin{equation}
\tau _{A_{i}|A_{j}|A_{k}}\left( \rho \right) \leq \frac{4\left\vert c\left(
a^{2}-d^{2}\right) \right\vert }{\left( \left\vert a\right\vert
^{2}+\left\vert d\right\vert ^{2}+2\left\vert c\right\vert ^{2}+1\right) ^{2}%
}\left( \frac{\left\vert \left\vert c\left( a^{2}-d^{2}\right) \right\vert
-\left\vert \left( a^{2}-c^{2}\right) \left( d^{2}-c^{2}\right) \right\vert
\right\vert }{\left\vert c\left( a^{2}-d^{2}\right) \right\vert +\left\vert
\left( a^{2}-c^{2}\right) \left( d^{2}-c^{2}\right) \right\vert }\right) .
\end{equation}%
Correct bound calculated in \cite{regu16} for this class of states is $\frac{%
4\left\vert c\left( a^{2}-d^{2}\right) \right\vert }{\left( \left\vert
a\right\vert ^{2}+\left\vert d\right\vert ^{2}+2\left\vert c\right\vert
^{2}+1\right) ^{2}}$.

\subsection{Class III}

Three tangle vanishes on reduced state obtained by tracing out qubit $A_{2}$
or $A_{4}$ from state 
\begin{equation}
\left\vert G_{ab}^{(3)}\right\rangle =a\left( \left\vert 0000\right\rangle
+\left\vert 1111\right\rangle \right) +b\left( \left\vert 0101\right\rangle
+\left\vert 1010\right\rangle \right) +\left\vert 0011\right\rangle
+\left\vert 0110\right\rangle ,
\end{equation}%
because $N_{4,8}^{A_{1}A_{3}A_{4}}=N_{4,8}^{A_{1}A_{2}A_{3}}=2\left\vert
I_{4,8}\right\vert $. On the other hand, if qubit $A_{3}$ is traced out then
non-zero three-qubit invariants $\left( I_{3}\right) _{A_{3}}^{0,4}=\frac{%
-4ab}{\left( 2\left\vert a\right\vert ^{2}+2\left\vert b\right\vert
^{2}+2\right) ^{2}}$ and $\left( I_{3}\right) _{A_{3}}^{2,2}=\frac{2\left(
a^{2}-b^{2}\right) ^{2}}{3\left( 2\left\vert a\right\vert ^{2}+2\left\vert
b\right\vert ^{2}+2\right) ^{2}}$, determine the three tangle. Four qubit
invariant $\left\vert I_{4,8}\right\vert =3\left\vert \left( I_{3}\right)
_{A_{l}}^{2,2}\right\vert ^{2}$ and pure state three-way correlation are
found to be $16\left( N_{4,8}^{A_{1}A_{2}A_{4}}-2\left\vert
I_{4,8}\right\vert \right) =16\left\vert \left( I_{3}\right)
_{A_{3}}^{0,4}\right\vert ^{2}$. Using the result given in Eq. (\ref{tauup5}%
), upper bound on three tangle for the partition $A_{1}A_{2}A_{4}$ reads as 
\begin{equation}
\tau _{A_{1}|A_{2}|A_{4}}\left( \rho \right) \leq \frac{4\left\vert
ab\right\vert }{\left( \left\vert a\right\vert ^{2}+\left\vert b\right\vert
^{2}+1\right) ^{2}}\frac{\left\vert 4\left\vert ab\right\vert -\left\vert
a^{2}-b^{2}\right\vert ^{2}\right\vert }{\left( 4\left\vert ab\right\vert
+\left\vert a^{2}-b^{2}\right\vert ^{2}\right) }.
\end{equation}%
Upper bound calculated in ref. \cite{regu14} is $\tau
_{A_{1}|A_{2}|A_{4}}\left( \rho \right) \leq \frac{4\left\vert ab\right\vert 
}{\left( \left\vert a\right\vert ^{2}+\left\vert b\right\vert ^{2}+1\right)
^{2}}$. Our upper bound may also be compared with the convex roof for the
same state reported in \cite{oste16} to be 
\begin{equation}
\left[ \tau _{A_{1}|A_{2}|A_{4}}\left( \rho \right) \right] ^{\frac{1}{2}%
}=\max \left( 0,\left( \frac{2\sqrt{\left\vert ab\right\vert }}{\left(
\left\vert a\right\vert ^{2}+\left\vert b\right\vert ^{2}+1\right) }\right) 
\frac{\left( 4\left\vert ab\right\vert -\left\vert a^{2}-b^{2}\right\vert
^{2}\right) }{4\left\vert ab\right\vert }\right) \text{.}
\end{equation}

\subsection{Class IV}

For entanglement class represented by 
\begin{align}
\left\vert G_{ab}^{\left( 4\right) }\right\rangle & =a\left( \left\vert
0000\right\rangle +\left\vert 1111\right\rangle \right) +\frac{a+b}{2}\left(
\left\vert 1010\right\rangle +\left\vert 0101\right\rangle \right)   \notag
\\
& +\frac{a-b}{2}\left( \left\vert 0110\right\rangle +\left\vert
1001\right\rangle \right) +\frac{i}{\sqrt{2}}\left( \left\vert
0010\right\rangle +\left\vert 0001\right\rangle +\left\vert
0111\right\rangle +\left\vert 1011\right\rangle \right) 
\end{align}%
all four reduced density matrices are found to have the same upper bound on
three-tangle. Taking up the case of qubits $A_{1}A_{2}A_{3}$, only non-zero
pure-state three-qubit invariant is $\left( I_{3}\right) _{A_{4}}^{0,4}$.
Since $I_{4,8}=0$, three-tangle is equal to 
\begin{equation}
\tau _{A_{1}|A_{2}|A_{3}}\left( \rho \right) =\sqrt{N_{4,8}^{A_{1}A_{2}A_{3}}%
}=4\left\vert \left( I_{3}\right) _{A_{4}}^{0,4}\right\vert .  \label{t123}
\end{equation}%
Substituting the value of $\left( I_{3}\right) _{A_{4}}^{0,4}$ in Eq. (\ref%
{t123}) and using $\tau _{A_{1}|A_{2}|A_{3}}\left( \rho \right) =\tau
_{A_{i}|A_{j}|A_{k}}\left( \rho \right) $, we have the inequality 
\begin{equation}
\tau _{A_{i}|A_{j}|A_{k}}\left( \rho \right) \leq \frac{2\left\vert
a^{2}-b^{2}\right\vert }{\left( 2+3\left\vert a\right\vert ^{2}+\left\vert
b\right\vert ^{2}\right) ^{2}}.
\end{equation}%
which is the same as reported for this state in \cite{regu14}.

\subsection{Class V}

For \ representative of entanglement class V, which reads as 
\begin{align}
\left\vert G_{a}^{(5)}\right\rangle & =a\left( \left\vert 0000\right\rangle
+\left\vert 1111\right\rangle +\left\vert 0101\right\rangle +\left\vert
1010\right\rangle \right)  \notag \\
& +i\left\vert 0001\right\rangle +\left\vert 0110\right\rangle -i\left\vert
1011\right\rangle ,
\end{align}%
non-zero three-qubit invariants of interest are 
\begin{equation}
\left( I_{3}\right) _{A_{4}}^{0,4}=\left( I_{3}\right) _{A_{2}}^{4,0}=\frac{%
-4a^{2}}{\left( 3+4\left\vert a\right\vert ^{2}\right) ^{2}},
\end{equation}%
and%
\begin{equation}
\left( I_{3}\right) _{A_{3}}^{1,3}=\frac{-2ia^{2}}{\left( 3+4\left\vert
a\right\vert ^{2}\right) ^{2}};\quad \left( I_{3}\right) _{A_{3}}^{0,4}=%
\frac{-1}{\left( 3+4\left\vert a\right\vert ^{2}\right) ^{2}}.
\end{equation}%
Consequently, $\tau _{A_{1}|A_{2}|A_{3}}\left( \rho \right) =\tau
_{A_{1}|A_{3}|A_{4}}\left( \rho \right) $, such that%
\begin{equation*}
\tau _{A_{1}|A_{2}|A_{3}}\left( \rho \right) \leq \frac{16\left\vert
a^{2}\right\vert }{\left( 3+4\left\vert a\right\vert ^{2}\right) ^{2}}.
\end{equation*}%
This bound coincides with the results from ref. \cite{oste16} and ref. \cite%
{regu14}.

For the marginal state obtained by tracing out qubit $A_{3}$, the values of $%
\left( I_{3}\right) _{A_{3}}^{0,4}$ and $\left( I_{3}\right) _{A_{3}}^{1,3}$
are substituted in Eq. (\ref{tauup6}) to obtain%
\begin{equation}
\tau _{A_{1}|A_{2}|A_{4}}\left( \rho \right) \leq \frac{4}{\left(
3+4\left\vert a\right\vert ^{2}\right) ^{2}}\left( \frac{1}{1+64\left\vert
a\right\vert ^{4}}\right) .
\end{equation}%
In comparison, upper bound on three tangle calculated by Osterloh (Eq. 37 of
ref. (\cite{oste16}) ) corresponds to%
\begin{equation}
\tau _{A_{1}|A_{2}|A_{4}}\left( \rho \right) \leq \frac{4}{\left(
3+4a^{2}\right) ^{2}}\left( \frac{1+64\left\vert a\right\vert ^{2}}{\left(
1+64\left\vert a\right\vert ^{4}\right) ^{2}}\right) \text{.}
\end{equation}

\subsection{Classes VI, VII, VIII and IX}

Non-zero pure state three-qubit invariants for class six state%
\begin{equation}
\left\vert G_{a}^{\left( 6\right) }\right\rangle =a\left( \left\vert
0000\right\rangle +\left\vert 1111\right\rangle \right) +\left\vert
0011\right\rangle +\left\vert 0101\right\rangle +\left\vert
0110\right\rangle ,
\end{equation}%
are given by $\left( I_{3}\right) _{A_{4}}^{2,2}=\left( I_{3}\right)
_{A_{3}}^{2,2}=\left( I_{3}\right) _{A_{2}}^{2,2}=\frac{a^{4}}{6\left(
3+2a^{2}\right) ^{2}}$. Consequently, $\tau _{A_{i}|A_{j}|A_{k}}\left( \rho
\right) $ is zero on states of the entanglement type represented by state $%
\left\vert G_{a}^{\left( 6\right) }\right\rangle $. States represented by 
\begin{equation}
\left\vert G^{\left( 7\right) }\right\rangle =\left\vert 0000\right\rangle
+\left\vert 0101\right\rangle +\left\vert 1000\right\rangle +\left\vert
1110\right\rangle ,
\end{equation}%
and 
\begin{equation}
\left\vert G_{ab}^{\left( 8\right) }\right\rangle =\left\vert
0000\right\rangle +\left\vert 1011\right\rangle +\left\vert
1101\right\rangle +\left\vert 1110\right\rangle ,
\end{equation}%
differ in the amount of two-way correlations. For both the states, $\tau
_{A_{1}|A_{j}|A_{k}}\left( \rho \right) =\frac{1}{4},\tau
_{A_{2}|A_{3}|A_{4}}\left( \rho \right) =0$, while state nine which reads as%
\begin{equation}
\left\vert G_{ab}^{\left( 9\right) }\right\rangle =\left\vert
0000\right\rangle +\left\vert 0111\right\rangle ,
\end{equation}%
has obviously $\tau _{A_{2}|A_{3}|A_{4}}\left( \rho \right) =\frac{1}{4}$.

\section{Upper bound on three tangle and optimal decomposition of a rank two
mixed state}

The procedure of section III may be used as an additional tool to find the
optimal decomposition $\{p_{i},\left\vert \phi _{i}\right\rangle $\} that
realizes the minimum in the definition (Eq. (\ref{mix3tangle})) of three
tangle of a mixed three-qubit state $\rho _{3}=\sum_{i}p_{i}\left\vert \phi
_{i}\right\rangle \left\langle \phi _{i}\right\vert $. To write down the
equations corresponding to Eq. (\ref{i40}) and Eq. (\ref{i04}), one
calculates relevant three-qubit invariants of the purification of the state.
Minimization may require solving a quartic equation. If an analytical
solution is not available, then the roots of the equation may be found
numerically. Consider a mixture of three-qubit pure states, 
\begin{equation}
\rho _{3}=p\left\vert \phi ^{\left( 0\right) }\right\rangle \left\langle
\phi ^{\left( 0\right) }\right\vert +\left( 1-p\right) \left\vert \phi
^{\left( 1\right) }\right\rangle \left\langle \phi ^{\left( 1\right)
}\right\vert ,  \label{ro123}
\end{equation}%
where $\left\langle \phi _{1}\right. \left\vert \phi _{0}\right\rangle =0$.
Purification of the state can be written as 
\begin{equation}
\left\vert \Psi _{4}\right\rangle =\sqrt{p}\left\vert \phi ^{\left( 0\right)
}\right\rangle \left\vert 0\right\rangle +\exp \left( i\theta \right) \sqrt{%
1-p}\left\vert \phi ^{\left( 1\right) }\right\rangle \left\vert
1\right\rangle .
\end{equation}%
Action of $U(x)=\frac{1}{\sqrt{1+\left\vert x\right\vert ^{2}}}\left[ 
\begin{array}{cc}
1 & -x^{\ast } \\ 
x & 1%
\end{array}%
\right] ,$ on fourth qubit of $\left\vert \Psi _{4}\right\rangle $ leads to
state%
\begin{equation}
\left\vert \Psi _{4}\left( x\right) \right\rangle =\sqrt{p^{\left( 0\right)
}(x)}\left\vert \phi ^{\left( 0\right) }(x)\right\rangle \left\vert
0\right\rangle +\sqrt{p^{\left( 1\right) }(x)}\left\vert \phi ^{\left(
1\right) }(x)\right\rangle \left\vert 1\right\rangle  \label{psix}
\end{equation}%
where%
\begin{equation}
\left\vert \phi ^{\left( 0\right) }(x)\right\rangle =\frac{\sqrt{p}%
\left\vert \phi ^{\left( 0\right) }\right\rangle -x^{\ast }\exp \left(
i\theta \right) \sqrt{1-p}\left\vert \phi ^{\left( 1\right) }\right\rangle }{%
\sqrt{p+\left\vert x\right\vert ^{2}\left( 1-p\right) }};p^{\left( 0\right)
}\left( x\right) =\frac{p+\left\vert x\right\vert ^{2}\left( 1-p\right) }{%
1+\left\vert x\right\vert ^{2}},
\end{equation}%
and%
\begin{equation}
\left\vert \phi ^{\left( 1\right) }(x)\right\rangle =\frac{\exp \left(
i\theta \right) \sqrt{1-p}\left\vert \phi ^{\left( 1\right) }\right\rangle +x%
\sqrt{p}\left\vert \phi ^{\left( 0\right) }\right\rangle }{\sqrt{p\left\vert
x\right\vert ^{2}+\left( 1-p\right) }};p^{\left( 1\right) }\left( x\right)
=1-p^{\left( 0\right) }\left( x\right) .
\end{equation}%
The upper bound on three tangle of reduced state%
\begin{equation*}
\rho (x)=\sum_{i=0}^{1}p^{\left( i\right) }(x)\left\vert \phi ^{\left(
i\right) }(x)\right\rangle \left\langle \phi ^{\left( i\right)
}(x)\right\vert ,
\end{equation*}%
subject to the condition that $\rho (x)=\rho _{3}$, can be found by using
the procedure outlined in section II. However, proper analysis of three
tangles $\tau _{A_{1}|A_{2}|A_{3}}\left( \left\vert \phi ^{\left( 0\right)
}(x)\right\rangle \right) $ and $\tau _{A_{1}|A_{2}|A_{3}}\left( \left\vert
\phi ^{\left( 1\right) }(x)\right\rangle \right) $ along with the respective
vectors, further aids in improving on the upper bound.

To illustrate, we recover the results for a mixed state studied in ref. \cite%
{lohm06}, which reads as 
\begin{equation}
\rho _{3}=p\left\vert GHZ\right\rangle \left\langle GHZ\right\vert +\left(
1-p\right) \left\vert W\right\rangle \left\langle W\right\vert ,
\end{equation}%
such that $\left\vert GHZ\right\rangle =\frac{1}{\sqrt{2}}\left( \left\vert
000\right\rangle +\left\vert 111\right\rangle \right) $ and $\left\vert
W\right\rangle =\frac{1}{\sqrt{3}}\left( \left\vert 100\right\rangle
+\left\vert 010\right\rangle +\left\vert 001\right\rangle \right) $. Values
of relevant three-qubit invariants for the purification 
\begin{equation}
\left\vert \Psi \right\rangle =\sqrt{p}\left\vert GHZ\right\rangle
\left\vert 0\right\rangle +\exp \left( i\theta \right) \sqrt{\left(
1-p\right) }\left\vert W\right\rangle \left\vert 0\right\rangle ,
\end{equation}%
are $\left( I_{3}\right) _{A_{4}}^{4,0}=\frac{p^{2}}{4}$ and $4\left(
I_{3}\right) _{A_{4}}^{1,3}=4\frac{\exp \left( i3\theta \right) \sqrt{%
p\left( 1-p\right) ^{3}}}{3\sqrt{6}}$. State $\left\vert \Psi _{4}\left(
x\right) \right\rangle $ (Eq. (\ref{psix}))obtained after a unitary
transformation $U(x)$ on fourth qubit contains normalized three-qubit states%
\begin{equation}
\left\vert \phi ^{\left( 0\right) }(x,\theta )\right\rangle =\frac{\sqrt{p}%
\left\vert GHZ\right\rangle -x^{\ast }\exp \left( i\theta \right) \sqrt{1-p}%
\left\vert W\right\rangle }{\sqrt{p+\left\vert x\right\vert ^{2}\left(
1-p\right) }},\quad p^{\left( 0\right) }(x)=\frac{p+\left\vert x\right\vert
^{2}\left( 1-p\right) }{1+\left\vert x\right\vert ^{2}},
\end{equation}%
and%
\begin{equation*}
\left\vert \phi ^{\left( 1\right) }(x,\theta )\right\rangle =\frac{\exp
\left( i\theta \right) \sqrt{1-p}\left\vert W\right\rangle +x\sqrt{p}%
\left\vert GHZ\right\rangle }{\sqrt{\left( 1-p\right) +p\left\vert
x\right\vert ^{2}}};\quad p^{\left( 1\right) }(x)=1-p^{\left( 0\right) }(x),
\end{equation*}%
such that $x$ dependent three tangle is given by%
\begin{equation*}
\left[ \tau _{A_{1}|A_{2}|A_{3}}\left( \rho \left( x\right) \right) \right]
^{\frac{1}{2}}=\sum_{i=1,2}p^{\left( i\right) }(x)\left[ \tau
_{A_{1}|A_{2}|A_{3}}\left( \left\vert \phi ^{\left( i\right) }(x,\theta
)\right\rangle \right) \right] ^{\frac{1}{2}}\text{.}
\end{equation*}%
Three tangles of vectors in the superposition are 
\begin{equation*}
\tau _{A_{1}|A_{2}|A_{3}}\left( \left\vert \phi ^{\left( 0\right) }(x,\theta
)\right\rangle \right) =\frac{4\left\vert \left( I_{3}\right)
_{A_{4}}^{4,0}-4\left( x^{\ast }\right) ^{3}\left( I_{3}\right)
_{A_{4}}^{1,3}\right\vert }{\left( p+\left( 1-p\right) \left\vert
x\right\vert ^{2}\right) ^{2}},
\end{equation*}%
and%
\begin{equation*}
\tau _{A_{1}|A_{2}|A_{3}}\left( \left\vert \phi ^{\left( 1\right) }(x,\theta
)\right\rangle \right) =\frac{4\left\vert x^{4}\left( I_{3}\right)
_{A_{4}}^{4,0}+4x\left( I_{3}\right) _{A_{4}}^{1,3}\right\vert }{\left(
\left\vert x\right\vert ^{2}p+\left( 1-p\right) \right) ^{2}}.
\end{equation*}%
Substituting the values of $\left( I_{3}\right) _{A_{4}}^{4,0}$ and $4\left(
I_{3}\right) _{A_{4}}^{1,3}$ in $\tau _{A_{1}|A_{2}|A_{3}}\left( \left\vert
\phi ^{\left( 0\right) }(x,\theta )\right\rangle \right) $, we obtain 
\begin{equation*}
\tau _{A_{1}|A_{2}|A_{3}}\left( \left\vert \phi ^{\left( 0\right) }(x,\theta
)\right\rangle \right) =\frac{4\left\vert \frac{p^{2}}{4}-4\left( x^{\ast
}\right) ^{3}\frac{\exp \left( i3\theta \right) \sqrt{p\left( 1-p\right) ^{3}%
}}{3\sqrt{6}}\right\vert }{\left( p+\left( 1-p\right) \left\vert
x\right\vert ^{2}\right) ^{2}}
\end{equation*}%
which is a periodic function of $\theta $ with a period of $2\pi /3$. For
the choice $\theta _{n}=2\pi n/3$, $n=0$, $1$, $2$, three-tangle $\tau
_{A_{1}|A_{2}|A_{3}}\left( \left\vert \phi ^{\left( 0\right) }(x_{0},\theta
_{n})\right\rangle \right) $ becomes zero for $x_{0}=\left( \frac{3\times 2^{%
\frac{5}{3}}p}{16\left( 1-p\right) }\right) ^{\frac{1}{2}}$. A closer
examination shows that for $p\leq 0.626851$ the value of $x_{0}$ lies within
the range $0\leq \left\vert x\right\vert \leq 1$, while for $p>0.626851$, $%
\tau _{A_{1}|A_{2}|A_{3}}\left( \left\vert \phi ^{\left( 0\right) }(x,\theta
_{n})\right\rangle \right) >0$ being minimum at $x=1$. For $0\leq p\leq
0.626851$, three-tangle $\left[ \tau _{A_{1}|A_{2}|A_{3}}\left( \rho
_{3}\left( x_{0}\right) \right) \right] =0$, for the mixed state%
\begin{equation*}
\rho _{3}\left( x_{0}\right) =\frac{1}{3}\sum_{n=0}^{2}\left\vert \phi
^{\left( 0\right) }(x_{0},\theta _{n})\right\rangle \left\langle \phi
^{\left( 0\right) }(x_{0},\theta _{n})\right\vert ,
\end{equation*}%
where%
\begin{equation*}
\left\vert \phi ^{\left( 0\right) }(x_{0},\theta _{n})\right\rangle =\frac{%
4\left\vert GHZ\right\rangle -\exp \left( i\theta _{n}\right) \sqrt{3\times
2^{\frac{5}{3}}}\left\vert W\right\rangle }{4\sqrt{\left( 1+\frac{3}{8}2^{%
\frac{2}{3}}\right) }}.
\end{equation*}%
Hence the decomposition of $\rho _{3}$ of Eq. (\ref{ro123}), with $\left[
\tau _{A_{1}|A_{2}|A_{3}}\left( \rho _{3}\left( x_{0}\right) \right) \right]
=0$ can be written as%
\begin{equation*}
\rho _{3}=p\left( 1+\frac{3}{8}2^{\frac{2}{3}}\right) \rho _{3}\left(
x_{0}\right) +\left( 1-p\left( 1+\frac{3}{8}2^{\frac{2}{3}}\right) \right)
\left\vert W\right\rangle \left\langle W\right\vert \text{.}
\end{equation*}

Since for $p>0.626851,$ vectors $\left\vert \phi ^{\left( 0\right)
}(1,\theta _{n})\right\rangle $ have lowest value of three-tangle the upper
bound on three-tangle is given by%
\begin{equation*}
\tau _{A_{1}|A_{2}|A_{3}}\left( \rho _{3}\left( 1\right) \right) =\left\vert 
\frac{p^{2}}{4}-\frac{4\sqrt{p\left( 1-p\right) ^{3}}}{3\sqrt{6}}\right\vert 
\text{,}
\end{equation*}%
and the corresponding decomposition is 
\begin{equation*}
\rho _{3}=\frac{1}{3}\sum_{n=0}^{2}\left\vert \phi ^{\left( 0\right)
}(1,\theta _{n})\right\rangle \left\langle \phi ^{\left( 0\right) }(1,\theta
_{n})\right\vert .
\end{equation*}%
Similar arguments may be used to find upper bound on three-tangle of an
arbitrary rank-two mixed state of three qubits.

\section{Concluding remarks}

For most of the four-qubit states, our bound on three-tangle of reduced
state is tighter than that used in ref. \cite{regu14}. A careful examination
shows that the upper bounds on three tangles listed in Table I of ref. \cite%
{regu14} are given by $4\sqrt{N_{4,8}^{A_{i}A_{j}A_{k}}-\left\vert
2I_{4,8}\right\vert }$. On the other hand for states that correspond to
cases (vi), (v), and (vi) of section III, three tangle satisfies%
\begin{equation}
\tau _{A_{i}|A_{j}|A_{k}}\left( \rho \right) \leq 4F\sqrt{%
N_{4,8}^{A_{i}A_{j}A_{k}}-\left\vert 2I_{4,8}\right\vert },\quad F\leq 1.
\end{equation}%
In ref. \cite{oste16} unitary on three qubit states is used to obtain
minimal characteristic curves to construct convex roof of three tangle for
nine classes of four-qubit states. Comparing the upper bound obtained by
unitary on fourth qubit with results for tangle corresponding to minimal
characteristic curve in ref. \cite{oste16} it is noted that for states in
class II, class III, and class V our value is lower for certain ranges of
state parameters than that of ref. \cite{oste16}. For all other cases, the
result obtained is the same.

Correct understanding of relation between pure state correlations and
entanglement of marginal states is crucial to discovering the form of
monogamy relations for multipartite entanglement. After examining the upper
bounds for nine classes of four-qubit states $\left\vert \Psi \right\rangle $%
, we conclude that%
\begin{eqnarray}
4\left\vert \left( I_{3}\right) _{A_{4}}^{4,0}\right\vert +4\left\vert
\left( I_{3}\right) _{A_{4}}^{0,4}\right\vert  &\geq &\tau
_{A_{1}|A_{2}|A_{3}}^{up}\left( \rho _{123}\right) \geq \tau
_{A_{1}|A_{2}|A_{3}}\left( \rho _{123}\right) \text{,}  \label{upthree} \\
4\left\vert \left( I_{3}\right) _{A_{3}}^{4,0}\right\vert +4\left\vert
\left( I_{3}\right) _{A_{3}}^{0,4}\right\vert  &\geq &\tau
_{A_{1}|A_{2}|A_{4}}^{up}\left( \rho _{124}\right) \geq \tau
_{A_{1}|A_{2}|A_{4}}\left( \rho _{124}\right) \text{,}
\end{eqnarray}%
and%
\begin{equation}
4\left\vert \left( I_{3}\right) _{A_{2}}^{4,0}\right\vert +4\left\vert
\left( I_{3}\right) _{A_{2}}^{0,4}\right\vert \geq \tau
_{A_{1}|A_{3}|A_{4}}^{up}\left( \rho _{134}\right) \geq \tau
_{A_{1}|A_{3}|A_{4}}\left( \rho _{134}\right) \text{.}
\end{equation}%
This result is used in \cite{shar17} to analytically write down the correct
monogamy inequality for four-qubit states.

To conclude, the method outlined in this letter can be used to obtain three
tangle of a rank-two three-qubit mixed state. Three-tangles are of interest
to establish the connection between condensed-matter physics and quantum
information \cite{amic08} as well as to better understand the connection
between quantum correlations in spin systems undergoing quantum phase
transitions \cite{werl10}.

Financial support from Universidade Estadual de Londrina PR, Brazil is
acknowledged.

\appendix

\section{Degree four three qubit invariants of four-qubit state}

\label{A1}

Degree four three-qubit invariants of four-qubit state relevant to
constructing the upper bound on $\tau _{A_{1}|A_{2}|A_{3}}\left( \rho
\right) $ in terms of two-qubit invariants for the pair $A_{1}A_{2}$ are
listed below:%
\begin{equation}
\left( I_{3}\right) _{A_{4}}^{4,0}=\left( D_{\left( A_{4}\right)
_{0}}^{000}+D_{\left( A_{4}\right) _{0}}^{001}\right) ^{2}-4D_{\left(
A_{3}\right) _{1}\left( A_{4}\right) _{0}}^{00}D_{\left( A_{3}\right)
_{0}\left( A_{4}\right) _{0}}^{00},
\end{equation}%
\begin{eqnarray}
\left( I_{3}\right) _{A_{4}}^{3,1} &=&\frac{1}{2}\left( D_{\left(
A_{4}\right) _{0}}^{000}+D_{\left( A_{4}\right) _{0}}^{001}\right) \left(
D^{0000}+D^{0001}+D^{0010}+D^{0011}\right)  \notag \\
&&-\left[ D_{\left( A_{3}\right) _{1}\left( A_{4}\right) _{0}}^{00}\left(
D_{\left( A_{3}\right) _{0}}^{000}+D_{\left( A_{3}\right) _{0}}^{001}\right)
+D_{\left( A_{3}\right) _{0}\left( A_{4}\right) _{0}}^{00}\left( D_{\left(
A_{3}\right) _{1}}^{000}+D_{\left( A_{3}\right) _{1}}^{001}\right) \right] ,
\end{eqnarray}

\begin{eqnarray}
\left( I_{3}\right) _{A_{4}}^{2,2} &=&\frac{1}{6}\left(
D^{0000}+D^{0001}+D^{0010}+D^{0011}\right) ^{2}  \notag \\
&&-\frac{2}{3}\left( D_{\left( A_{3}\right) _{1}}^{000}+D_{\left(
A_{3}\right) _{1}}^{001}\right) \left( D_{\left( A_{3}\right)
_{0}}^{000}+D_{\left( A_{3}\right) _{0}}^{001}\right)  \notag \\
&&+\frac{1}{3}\left( D_{\left( A_{4}\right) _{0}}^{000}+D_{\left(
A_{4}\right) _{0}}^{001}\right) \left( D_{\left( A_{4}\right)
_{1}}^{000}+D_{\left( A_{4}\right) _{1}}^{001}\right)  \notag \\
&&-\frac{2}{3}\left( D_{\left( A_{3}\right) _{1}\left( A_{4}\right)
_{0}}^{00}D_{\left( A_{3}\right) _{0}\left( A_{4}\right)
_{1}}^{00}+D_{\left( A_{3}\right) _{0}\left( A_{4}\right)
_{0}}^{00}D_{\left( A_{3}\right) _{1}\left( A_{4}\right) _{1}}^{00}\right) ,
\end{eqnarray}%
\begin{eqnarray}
\left( I_{3}\right) _{A_{4}}^{1,3} &=&\frac{1}{2}\left(
D^{0000}+D^{0001}+D^{0010}+D^{0011}\right) \left( D_{\left( A_{4}\right)
_{1}}^{000}+D_{\left( A_{4}\right) _{1}}^{001}\right)  \notag \\
&&-\left[ D_{\left( A_{3}\right) _{1}\left( A_{4}\right) _{1}}^{00}\left(
D_{\left( A_{3}\right) _{0}}^{000}+D_{\left( A_{3}\right) _{0}}^{001}\right)
+\left( D_{\left( A_{3}\right) _{1}}^{000}+D_{\left( A_{3}\right)
_{1}}^{001}\right) D_{\left( A_{3}\right) _{0}\left( A_{4}\right) _{1}}^{00}%
\right] ,
\end{eqnarray}%
and%
\begin{equation}
\left( I_{3}\right) _{A_{4}}^{0,4}=\left( D_{\left( A_{4}\right)
_{1}}^{000}+D_{\left( A_{4}\right) _{1}}^{001}\right) ^{2}-4D_{\left(
A_{3}\right) _{1}\left( A_{4}\right) _{1}}^{00}D_{\left( A_{3}\right)
_{0}\left( A_{4}\right) _{1}}^{00}.
\end{equation}

\end{document}